\documentclass[12pt]{iopart}

\usepackage{graphicx}

\usepackage{iopams}
\usepackage{pslatex}
\usepackage{txfonts}

\newcommand{\sqrts}{\sqrt{s}}
\newcommand{\sqrtsnn}{\sqrt{s_{_{NN}}}}

\providecommand{\ups}{\Upsilon}

\providecommand{\PYTHIA}{{\sc pythia }}

\newcommand{\Et}{E_{\rm T}}                    
\def\ETa{E_{T,\,1}}
\def\ETb{E_{T,\,2}}

\providecommand{\gaga}{\gamma\,\gamma}

\providecommand{\gA}{\gamma\,A}
\providecommand{\gpb}{\gamma\,$Pb$}

\providecommand{\elel}{e^+e^-}
\providecommand{\mumu}{\mu^+\mu^-}
\providecommand{\lele}{l^{+}\,l^{-}}

\begin{document}

\title[Small-$x$ QCD studies with CMS at the LHC]{Small-$x$ QCD studies with CMS at the LHC}

\author{David d'Enterria for the CMS collaboration}
\address{CERN, PH-EP, CH-1211 Geneva 23}

\begin{abstract}
\hspace{-0.4cm}
The capabilities of the CMS experiment to study the low-$x$ parton structure and QCD 
evolution in the proton and the nucleus at LHC energies are presented through 
four different measurements, to be carried out in Pb-Pb at $\sqrtsnn$ = 5.5 TeV:
(i) the charged hadron rapidity density $dN_{\rm ch}/d\eta$ and (ii) the
ultraperipheral (photo)production of $\ups$; and in p-p at $\sqrts$~=~14~TeV: 
(iii) inclusive forward jets and (iv) Mueller-Navelet dijets 
(separated by $\Delta\eta\gtrsim$ 8).
\end{abstract}


%

\section*{Introduction}

At high energies, the cross-sections of all {\it hadronic objects} (protons, nuclei, or even photons 
``fluctuating'' 
into $q\bar{q}$ vector states) are dominated by scatterings involving 
gluons. Gluons clearly outnumber quarks in the small momentum fraction (low-$x$) range
of the parton distribution functions (PDFs) as a consequence of the 
QCD parton splitting probabilities described by the DGLAP~\cite{dglap} 
and BFKL~\cite{bfkl} evolution equations. The fast growth of the gluon densities $xG(x,Q^2)$ 
for decreasing $x$ conspicuously observed in DIS $ep$ at HERA~\cite{hera}, cannot however continue 
indefinitely since this would violate unitarity even for scatterings with $Q^2\gg\Lambda_{\rm QCD}^2$. 
For small enough $x$ values, gluons must start to recombine in a process known as gluon saturation~\cite{saturation}.
This phenomenon occurs when the size occupied by the partons becomes similar to the size of the hadron 
$\pi R^2$, or in terms of the {\it saturation momentum} $Q_s$ when: 
$Q^2\lesssim Q_s^2(x)\simeq \alpha_s\,xG(x,Q^2)/\,\pi R^2$. 
$Q_s$ grows with the number $A$ of nucleons in the ``target'', 
the collision energy  $\sqrts$, and the rapidity of the gluon $y=\ln(1/x)$, according to:
$Q_s^2\sim A^{1/3}\,x^{-0.3} \sim A^{1/3}(\sqrt{s})^{0.3} \sim A^{1/3}e^{0.3\,y}$. 
The $A$ dependence implies that, at equal energies, saturation effects will be enhanced 
by factors as large as $A^{1/3}\approx$ 6 in a heavy nucleus ($A$ = 208 for Pb) compared to protons.
Theoretically, the regime of low-$x$ QCD can be effectively described in the ``Color Glass Condensate'' 
(CGC) framework, where all gluon fusions and multiple scatterings are ``resummed'' 
into classical high-density gluon wavefunctions~\cite{cgc}. The corresponding evolution is given 
in this case by the BK/JIMWLK~\cite{bkjimwlk} {\it non-linear} equations.\\ 


\noindent
Experimentally, the most direct way to access the low-$x$ PDFs in hadronic collisions is by measuring
perturbative probes (heavy-$Q$, jets, high-$p_T$ hadrons, prompt $\gamma$, ...) at large $\sqrts$ and 
{\it forward} rapidities~\cite{dde_heralhc06}. For a $2\rightarrow 2$ parton scattering, the {\it minimum} $x$ 
probed in a process with a particle of momentum $p_T$ produced at pseudo-rapidity 
$\eta$, is $x_{2}^{min} = x_T\,e^{-\eta}/(2-x_T\,e^{\eta})$ where $x_T=2p_T/\sqrt{s}$.
Thus, $x_2^{min}$ decreases by a factor of $\sim$10 every 2 units of rapidity.
The experimental capabilities of the CMS experiment are extremely well adapted for the 
study of low-$x$ phenomena with proton and ion beams. The acceptance of the CMS/TOTEM 
system is the largest ever available in a collider, and the detector is designed to measure 
different particles with excellent momentum resolution~\cite{russell}: jets $(|\eta|<6.6)$, 
$\gamma$ and $e^\pm$ ($|\eta|<$ 3), muons ($|\eta|<$ 2.5), hadrons ($|\eta|<$ 6.6), 
plus neutrals in the Zero-Degree Calorimeters (ZDCs, $|\eta|>$ 8.3). 
We present a selection of four observables measurable in CMS which are sensitive to parton 
saturation effects in the proton and nucleus wave-functions at LHC energies. Other relevant 
measurements (e.g. forward Drell-Yan in p-p at 14 TeV) are discussed in~\cite{FWD_PTDR}.

%

\section{Measurements in PbPb collisions at $\sqrtsnn$ = 5.5 TeV}



\subsection*{(1) Charged hadron PbPb rapidity density: $dN_{ch}/d\eta$}

\begin{figure}[!Hhtb]
\includegraphics[width=7.5cm,height=7.cm]{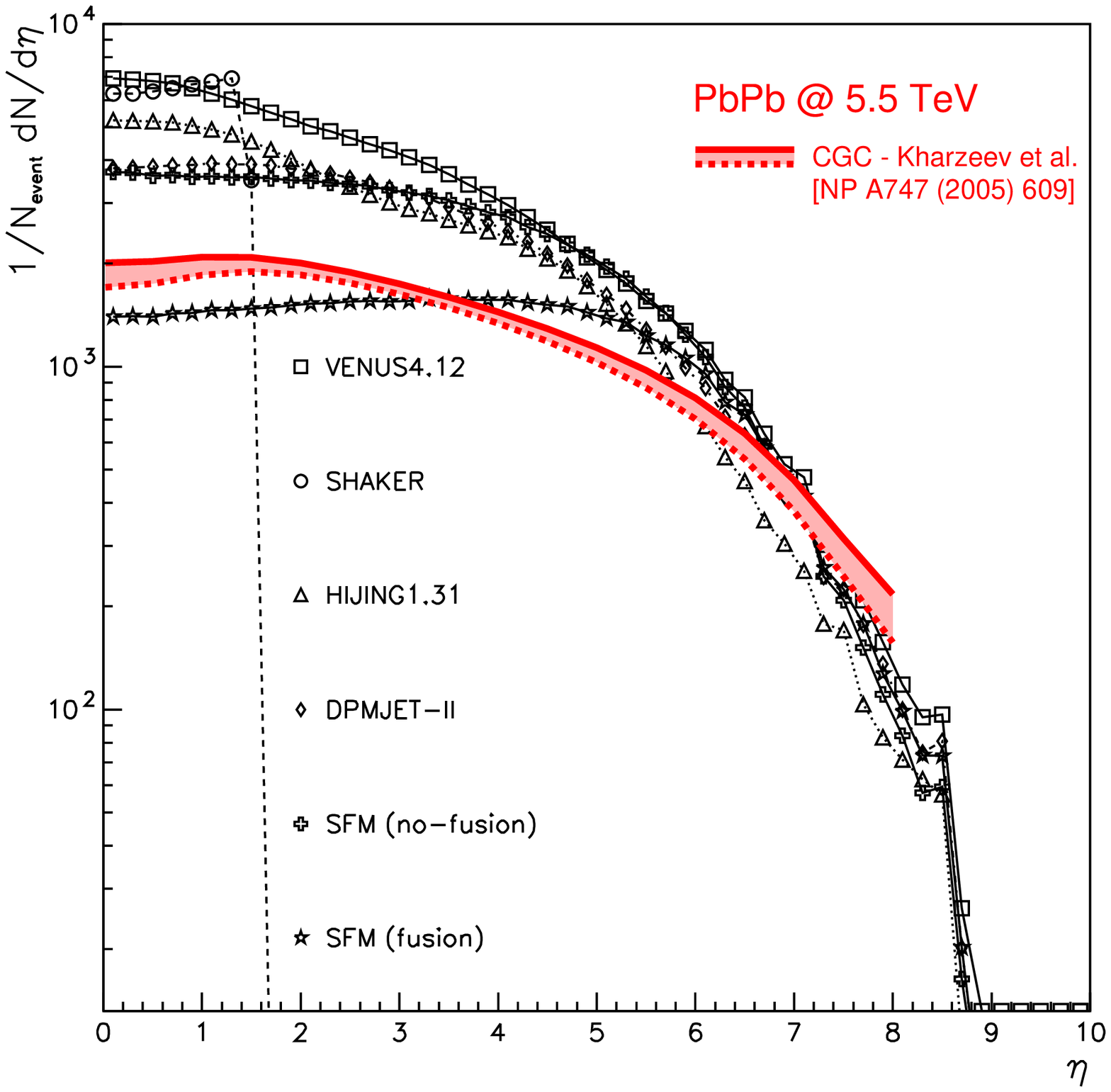}
\includegraphics[width=8.7cm,height=6.6cm]{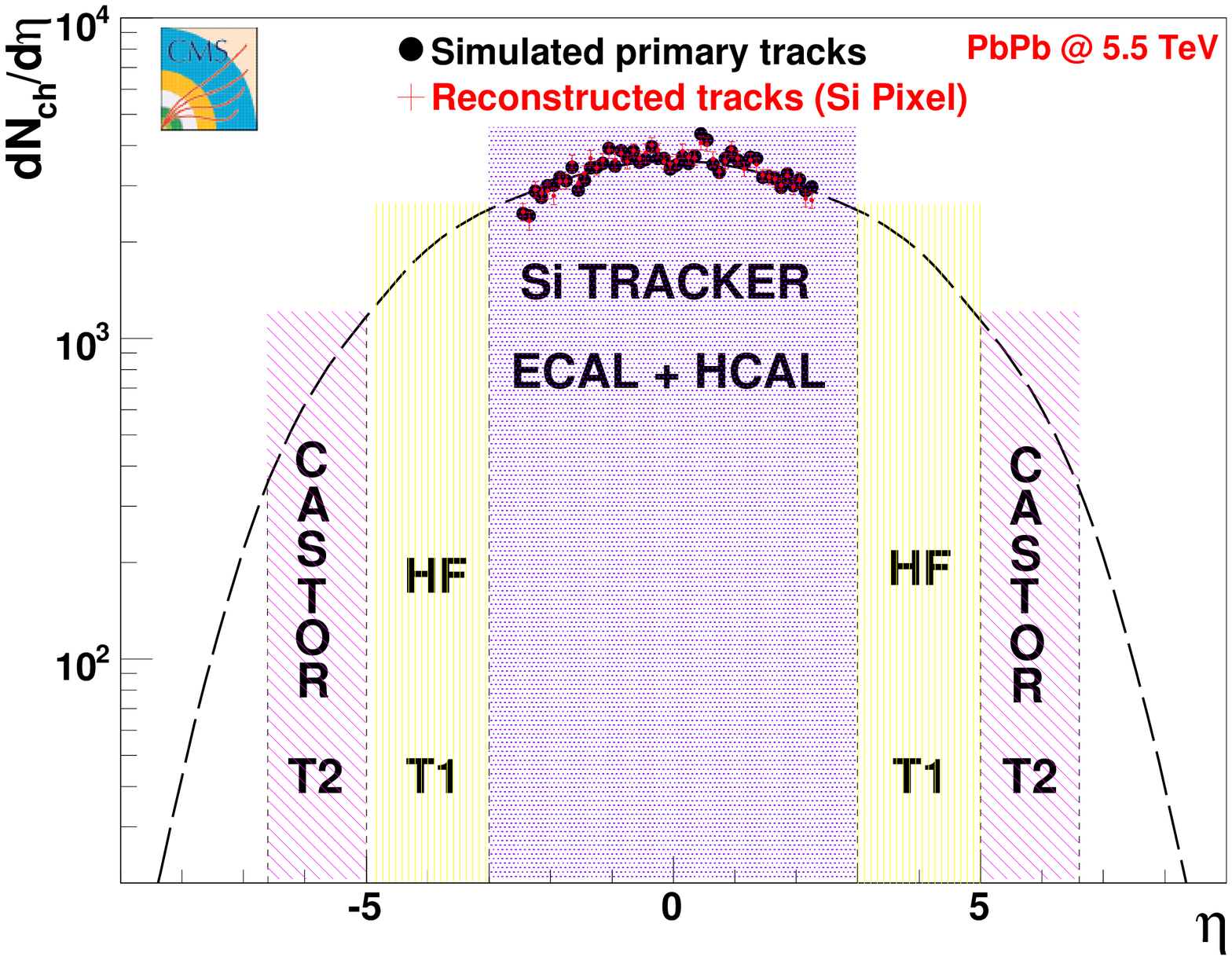}
\vskip -0.4cm
\caption{Left: Model predictions for $dN/d\eta$ in central PbPb at the LHC~\cite{kharzeev_kln,perco}.
Right: Range of particle rapidities covered by CMS (Si tracker, HF, CASTOR) 
and TOTEM (T1, T2 trackers). The {\sc hijing} PbPb distribution of primary simulated tracks 
within  $|\eta|<$ 2.5 (black dots) is compared to the reconstructed hits in the first layer of the 
Si tracker (red crosses)~\protect\cite{cms_hi_ptdr}.}
\label{fig:dNdeta}
\end{figure}

\noindent
In high-energy heavy-ion collisions, the \emph{hadron} rapidity density $dN/d\eta$ 
is directly related to the number of initially released \emph{partons} at a given $\eta$. CGC approaches
which effectively take into account a reduced initial parton flux 
in the nuclear PDFs, reproduce successfully the absolute hadron yields (as well as their centrality and 
$\sqrtsnn$ dependences) at SPS -- RHIC energies~\cite{kharzeev_kln,armesto04}. 
At LHC, the expected PbPb multiplicities are $dN/d\eta|_{\eta=0}\approx$~2000 
(Fig.~\ref{fig:dNdeta}, left). 
CMS simulation studies from hit counting in the innermost Si pixel layer ($|\eta|<$ 2.5) 
indicate that the occupancy remains less than 2\% and that, on an event-by-event basis, the 
reconstructed $dN_{\rm ch}/d\eta$ is within $\sim$2\% of the true primary multiplicity 
(Fig.~\ref{fig:dNdeta}, right)~\cite{cms_hi_ptdr}.
\vskip -0.4cm

\vspace{-0.5cm}
\subsection*{(2) $\ups$ photoproduction in ultra-peripheral 
PbPb $(\rightarrow \gpb) \rightarrow  \ups$ + Pb$^*$ Pb$^{(*)}$ collisions}

\begin{figure}[htb]
\includegraphics[width=7.7cm,height=6.5cm]{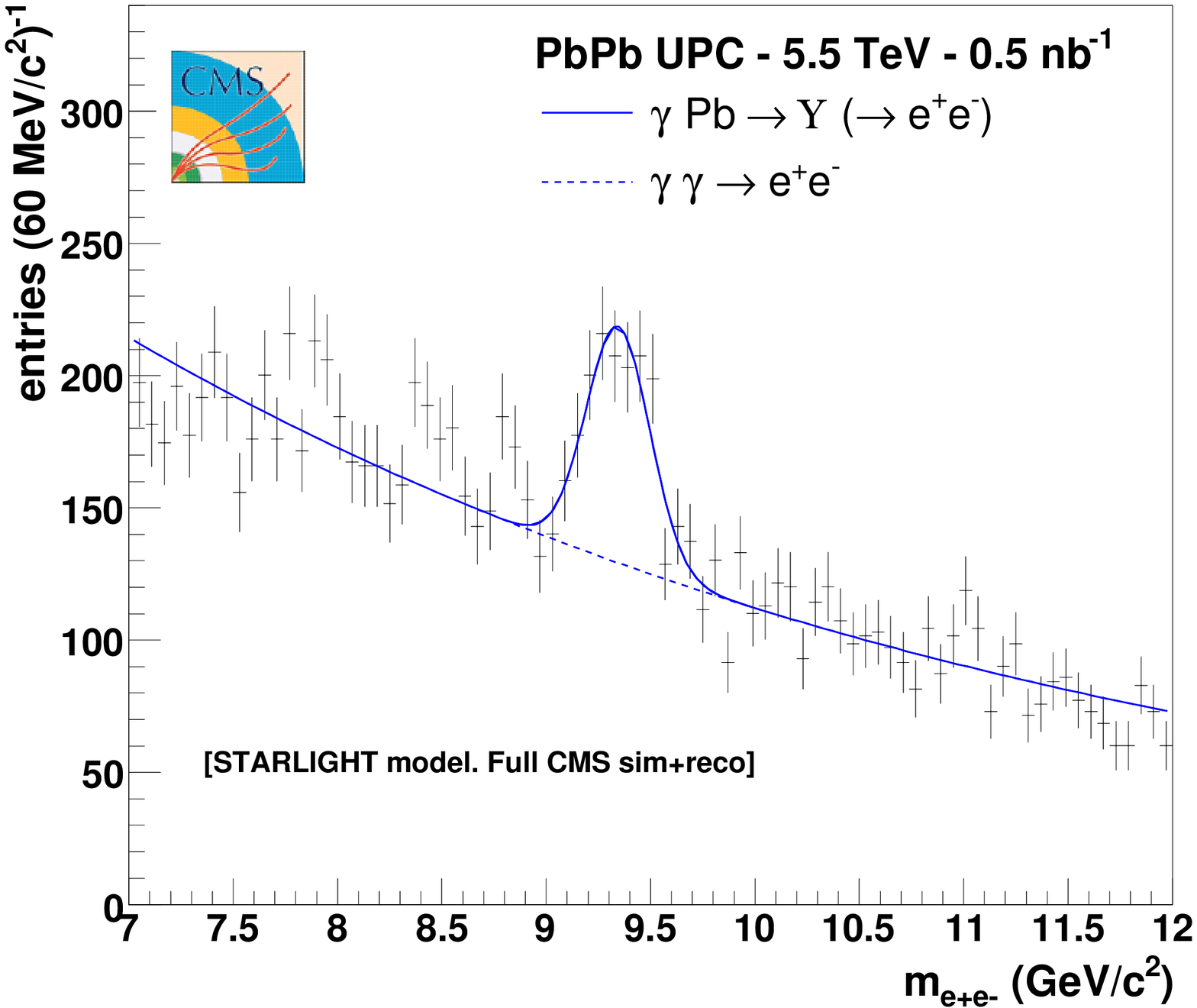}
\includegraphics[width=7.7cm,height=6.5cm]{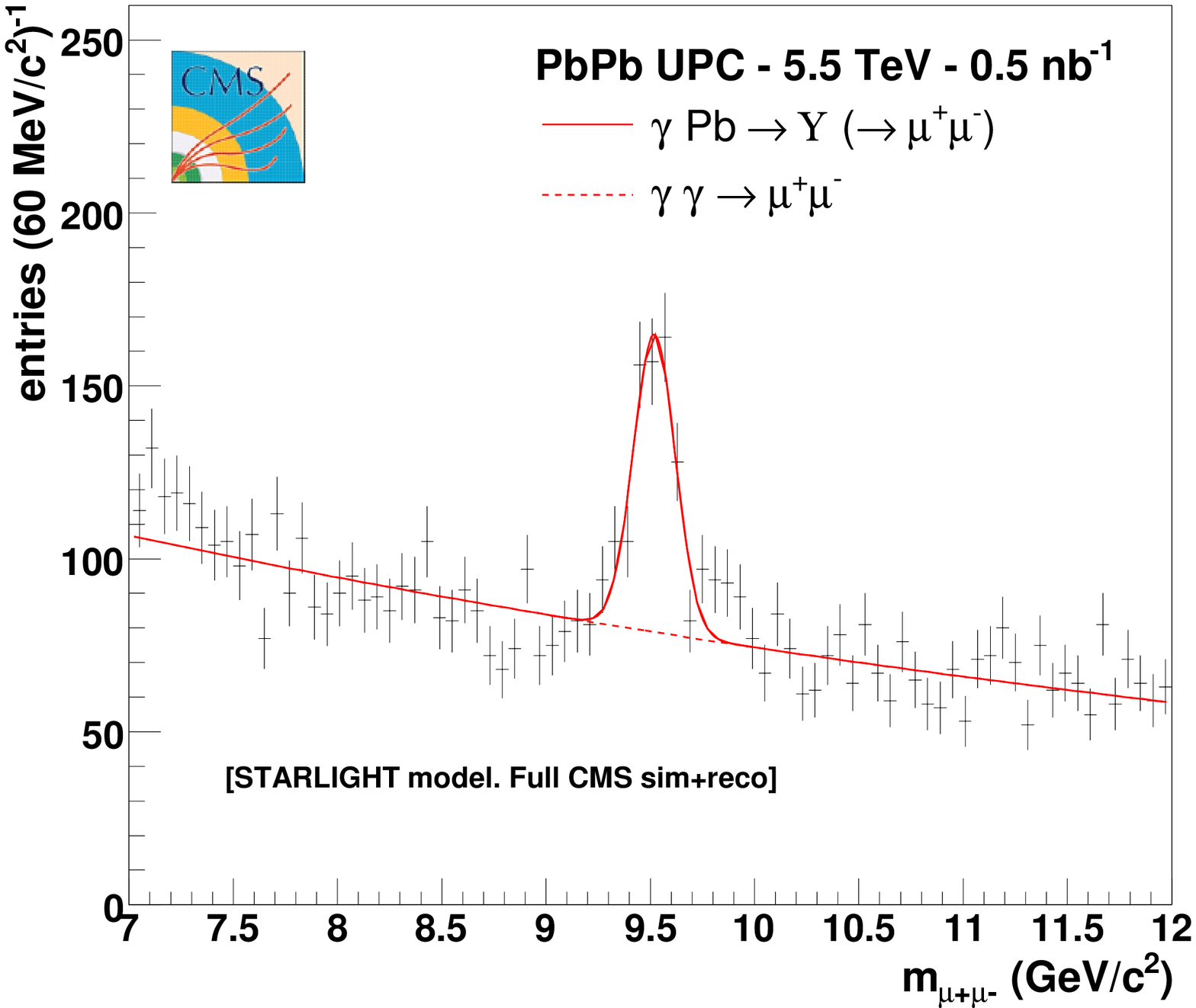}
\caption{Expected $\elel$ (left), $\mumu$ (right) invariant mass distributions from 
$\gamma\,$Pb$\rightarrow \Upsilon\,$Pb$^\star$ ($\ups\rightarrow \lele$, signal)
and $\gaga\rightarrow \lele$ (background) in UPC PbPb at $\sqrtsnn$ = 5.5 TeV in CMS.}
\label{fig:upc_ups_cms}
\end{figure}

\noindent
Ultraperipheral collisions (UPCs) of heavy ions generate strong electromagnetic fields 
(equivalent to a flux of quasi-real photons) which can be used to study
$xG(x,Q^2)$ via $Q\bar{Q}$ photoproduction~\cite{UPC_report}. 
Lead beams at 2.75 TeV have Lorentz factors $\gamma$ = 2930 leading to maximum 
photon energies $\omega_{\ensuremath{\it max}}\approx \gamma/R\sim$ 100 GeV
(for a nuclear radius $R=$ 6.5 fm)
and c.m. energies $W_{\gaga}^{\ensuremath{\it max}}\approx$ 160 GeV and $W^{\ensuremath{\it max}}_{\gA}\approx$ 1 TeV. 
The $x$ values probed in $\gpb\rightarrow\ups \,$Pb processes at $y$ = 2.5 can be as low as $x\sim 10^{-4}$. 
Full simulation+reconstruction~\cite{cms_hi_ptdr} of input distributions from 
the {\sc starlight} MC~\cite{starlight} show that CMS can measure $\ups\rightarrow \elel$, $\mumu$ 
within $|\eta|<$ 2.5, in UPCs tagged with neutrons detected in the ZDCs. Fig.~\ref{fig:upc_ups_cms} 
shows the reconstructed $dN/dm_{\lele}$ around the $\ups$ mass for 0.5 nb$^{-1}$ integrated PbPb
luminosity. With a total yield of $\sim$\,400 $\ups$, detailed 
$p_T$,$\eta$ studies can be carried out, to constrain the low-$x$ gluon 
density in the Pb nucleus.

%

\section{Measurements in pp collisions at $\sqrts$ =14 TeV}

\subsection*{(3) Inclusive forward jet production: pp $\rightarrow$ jet+X, with $3<|\eta_{\rm jet}|<5$}

\begin{figure}[htb]
\includegraphics[width=7.5cm,height=6.8cm]{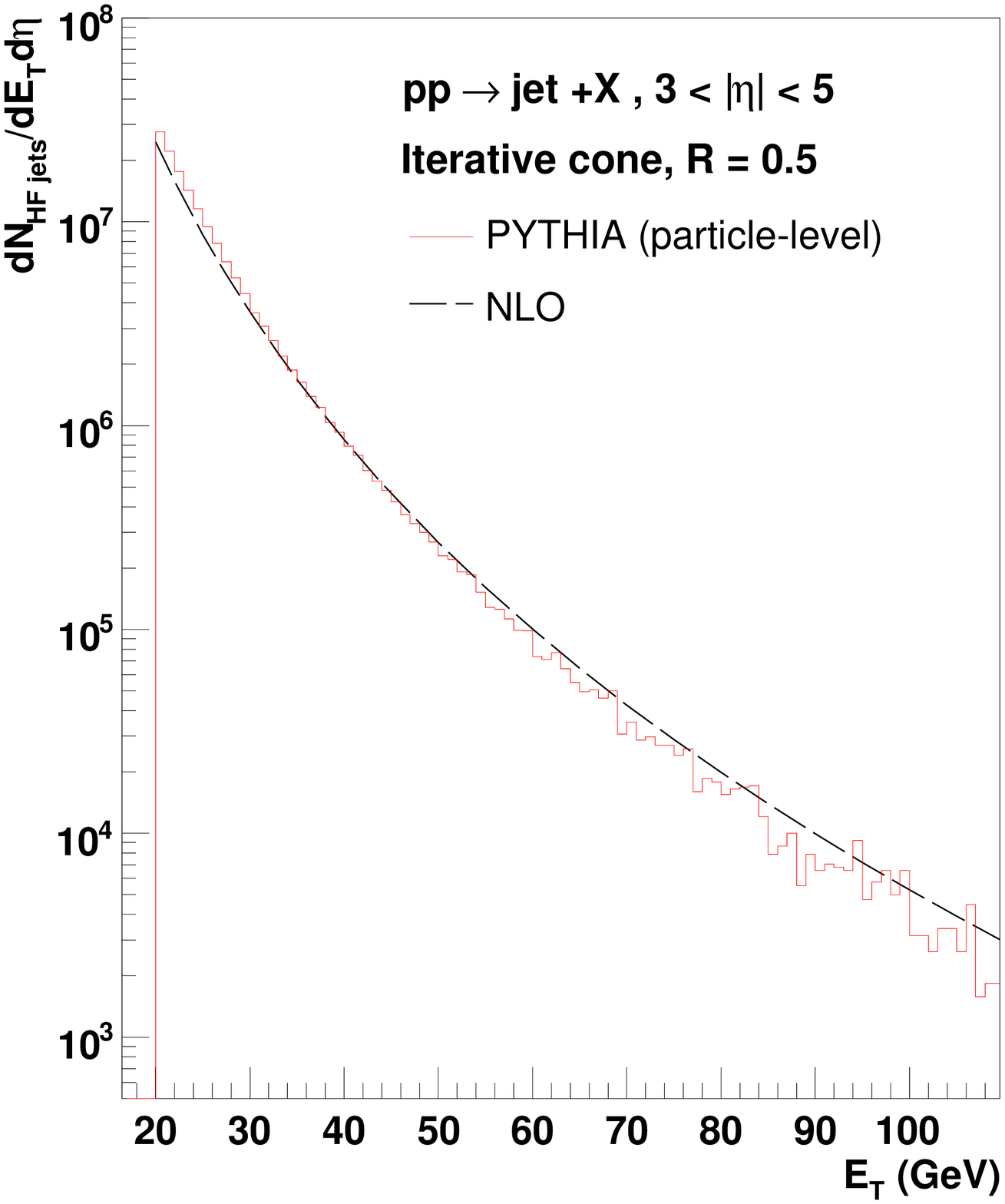}
\includegraphics[width=7.5cm,height=6.8cm]{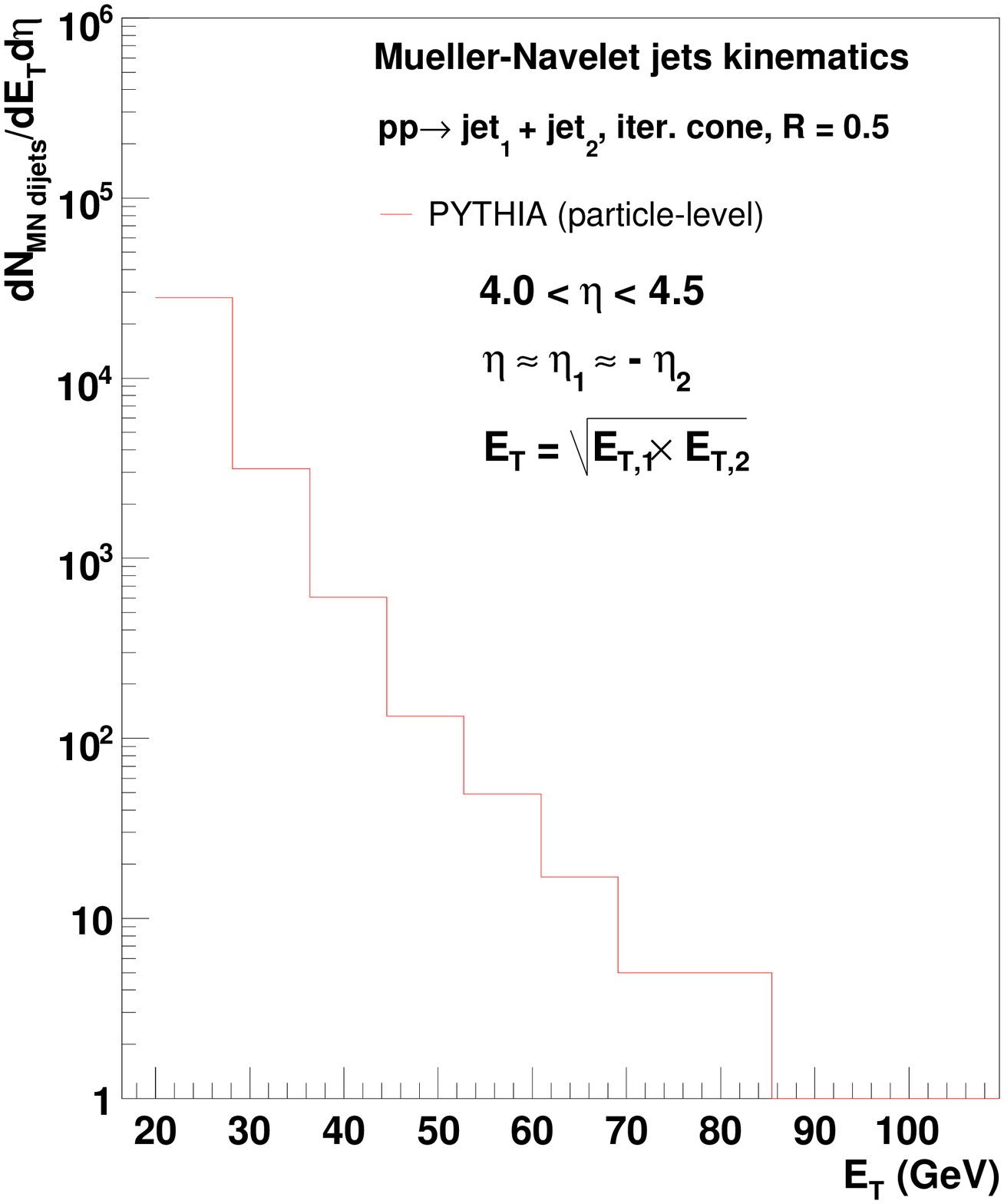}
\caption{Expected jet yields in pp at $\sqrts$ = 14 TeV (1 pb$^{-1}$) 
obtained from \PYTHIA 6.403 at the particle-level 
({\it no} full detector response, underlying-event or hadronization corrections).
Left: Single inclusive jets in HF ($3<|\eta|<5$) (compared to a NLO 
calculation with scale $\mu$ = $\Et$~\protect\cite{Jager:2004jh}).
Right: Dijets separated by $\Delta\eta =$ 8--9 with the M\"uller-Navelet kinematics cuts 
described in the text, as a function of  $E_T\equiv \sqrt{E_{T\,,1} \times E_{T\,,2}}$.}
\label{fig:pp_fwd_jets}
\end{figure}

Jet measurements at Tevatron have provided valuable information on the proton PDFs. 
At 14 TeV, the production of jets with $E_T\approx$ 20--100 GeV in the
CMS forward calorimeters (HF and CASTOR) probes the PDFs down to 
$x_{2}\approx 10^{-6}$~\cite{dde_heralhc06}. Figure~\ref{fig:pp_fwd_jets}-left shows the 
single inclusive jet spectrum in both HFs (3$<|\eta|<$5) expected for a short first run 
with just 1 pb$^{-1}$ integrated luminosity.
The spectrum has been obtained from a preliminary study using \PYTHIA 6.403 
with jet reconstruction at the {\it particle-level} (i.e.\ {\it no} detector effects are included 
apart from the HF tower $\eta-\phi$ granularity)~\cite{FWD_PTDR}. Although 
at such low $\Et$'s systematic uncertainties can be as large as $\sim$30\%, the available 
statistics for this study is very high.

\subsection*{(4) Mueller-Navelet dijets: pp $\rightarrow$ jet$_1$+jet$_2$, with large $\Delta\eta=\eta_2-\eta_1$ }

Inclusive dijet production at large pseudorapidity intervals -- M\"uller-Navelet (MN) jets -- 
has been considered an excellent testing ground for BFKL~\cite{mueller_navelet} 
and non-linear QCD~\cite{marquet05} evolutions. The large rapidity separation between partons 
enhances the available longitudinal momentum phase space for BFKL radiation. Gluon saturation 
effects are expected to reduce the (pure BFKL) MN cross section by a factor of $\sim$2 
for jets separated by $\Delta\eta\approx$ 9~\cite{marquet05}. 
In order to estimate the expected statistics for a short run without pile-up (1 pb$^{-1}$), 
we have selected the \PYTHIA events 
which pass the MN kinematics cuts: $|\ETa - \ETb| < 2.5$ GeV, $|\eta_1| - |\eta_2| < 0.25$, and
$\Delta\eta$ = 6 -- 10~\cite{FWD_PTDR}. Figure~\ref{fig:pp_fwd_jets}-right shows the results 
for $\Delta\eta$ = 8--9. The expected dijet yields for this $\eta$ separation indicate that 
these studies are clearly statistically feasible at the LHC.\\
\vspace*{-0.3cm}

\noindent
{\bf Acknowledgments.} Supported by 6th EU Framework Programme MEIF-CT-2005-025073.\\ 
\vspace*{-0.6cm}

\section*{References}

\end{document}